# Gating orbital memory with an atomic donor

Elze J. Knol[1], Brian Kiraly[1], Alexander N. Rudenko[1,], Werner M. J. van Weerdenburg[1], Mikhail I. Katsnelson[1,], Alexander A. Khajetoorians[1]

1. *Institute for Molecules and Materials, Radboud University, Nijmegen 6525AJ, The Netherlands*

**Orbital memory is defined by two stable valencies that can be electrically switched and read-out. To explore the influence of an electric field on orbital memory, we studied the distance-dependent influence of an atomic Cu donor on the state favorability of an individual Co atom on black phosphorus. Using low temperature scanning tunneling microscopy/spectroscopy, we characterized the electronic properties of individual Cu donors, corroborating this behavior with *ab initio* calculations based on density functional theory. We studied the influence of an individual donor on the charging energy and stochastic behavior of an individual Co atom. We found a strong impact on the state favorability in the stochastic limit. These findings provide quantitative information about the influence of local electric fields on atomic orbital memory.**

A single magnetic atom on a surface can exhibit multiple valencies, as predicted for various 3$d$ transition metal atoms on the surface of graphene [1, 2]. This concept was experimentally demonstrated, using scanning tunneling microscopy/spectroscopy (STM/STS), with individual Co atoms on the surface of semiconducting black phosphorus (BP) [3]. In this study, the two stable valencies of an individual Co atom residing in a hollow site were observed via their different charge densities, and could be switched electrically, serving as a so-called orbital memory. Moreover, the electric field generated by the probe could be used to ionize an individual Co atom, leading to stochastic fluctuations between the stable configurations characterized by telegraph noise in the tunneling current. It was later shown that the stochastic behavior of arrays of coupled orbital memories exhibits tunable multiplicity, a precursor to glassy dynamics characteristic in multi-well systems [4-6]. In this way, arrays of orbital memories are a promising platform to mimic machine learning at the atomic scale, due to their long-range connectivity and competing interactions.

Understanding the influence of an electric field on orbital memory is vital both fundamentally as well as for its utility, in analogy to the effect of a magnetic field on spin-based memory [7-9]. One way to quantify the role of an electric field on the bistable valency is to place well-defined atomic-scale dopants near individual orbital memories and probe the response of the bistable valency. In this vain, it has been shown that alkali atoms are charge donors on the surface of BP. For instance, it was shown that K atoms can be used to $n$-dope BP [10-12] and probe the anisotropic dielectric screening of the material [13]. However, K atoms are relatively challenging to use for STM-based studies, as they easily diffuse on the surface of BP and can be laterally perturbed by the tip-generated electric field.

Here, we explore the influence of a single donor, derived from an individual Cu atom, on the bistable valency of an isolated Co atom on BP. Using first principle calculations based on density functional theory (DFT), we find that individual Cu atoms residing in various binding sites as well as hydrogenated species can be modeled as individual positively charged ions on the surface. Based on these findings, we measured the changes in measured ionization energy of one of the orbital states of individual Co atoms, in proximity of a positively charged Cu donor, using STM/STS with atomic manipulation. Considering the various sources of electrostatic interactions, we show that the changes in ionization energy originate from local band bending arising from a screened Coulomb interaction of the Cu donor. We additionally observe a strong change in the orbital state-dependent lifetime of the ionized Co atom, in the stochastic limit, depending on the separation between a Co atom and Cu donor. This strong orbital state dependence in the measured lifetimes results from the interplay between the influence of the Cu gating field and the different dielectric screening of the two orbital states. We further find, from first principles calculations based on DFT, that each valency exhibits a significant electric dipole moment oriented perpendicular to the surface, leading to a substantially different response to the local electric field generated by the STM tip.

Fig. 1(a) illustrates an STM image of a cleaved BP surface after deposition of Co and Cu atoms [14-27]. In agreement with previous experiments [28], intrinsic vacancies are identified as elongated dumbbell-shaped protrusions. An individual Co atom preferentially adsorbs onto a top site, and its charge density can be identified as a bi-lobed, butterfly-like shape. The Co atom can be manipulated into a hollow site, using the STM tip [3]. In constant-current STM imaging, Co atoms residing in a hollow site can exhibit two different charge densities ($Co_{high}$ and $Co_{low}$). These two unique charge densities result from the bistable valency of the Co atom [3], and can be reversibly switched with a voltage pulse. Cu atoms can be differentiated from Co atoms due to their distinguishable charge densities in constant-current imaging. Three different Cu species were observed, all characterized by an ellipse-shaped outer depression whereas the internal pattern is distinct between all three species (Fig. 1(b-d) / Supplementary Information).

In order to quantify the electronic properties of each Cu species and relate this to the experimental observations, we performed DFT calculations for a Cu atom on the BP surface [14]. We calculated the band structure of Cu atoms residing on a top site ($Cu_T$) and in a hollow site ($Cu_H$) on BP, as well as of hydrogenated Cu in the hollow site ($CuH_H$) (Supplementary Fig. 3). In case of $Cu_H$ and $CuH_H$, the system is nonmagnetic, whereas a small magnetic moment of ~0.5 $\mu_B$ appears for $Cu_T$. We also calculated the spatial distribution of the charge densities projected on the valence states of BP (Fig. 1(e-g)), which can be directly associated with the STM images (Fig. 1(b-d)). Based on these calculations, we conclude that the experimentally observed species in Fig. 1(b) and Fig. 1(d) are the hollow and top site, respectively. We also confirm that the charge density observed in Fig. 1(c) coincides with CuH$_H$. From DFT calculations follows that Cu$_H$ / CuH$_H$ / Cu$_T$ species donate 0.7 / 0.12 / 0.24 electrons to the BP substrate, respectively. The electrons are mostly donated from the 4$s$ shell of Cu, whereas the 3$d$ shell remains fully occupied. These calculations are consistent with experimental observations of $n$-doping at higher Cu coverages (Supplementary Fig. 2). Therefore, we consider all Cu species as donors in the subsequent discussion, in line with previous literature [29-32]. In the following experiments, a relatively low areal density of $n_{Cu}$ = 0.165 x 10$^{12}$ cm$^{-2}$ was used to ensure minimal band shifts in the BP (< 8 meV).

In order to probe the influence of Cu donors on the bistable valency of individual Co atoms, we first studied the response of the Co$_{low}$ atom charging peak as a function of distance ($r$) from a single, isolated Cu atom. As shown in ref. [3], Co impurity states near $E_F$ can be pulled above $E_F$, via tip-induced band bending [33-35]. This leads to a peak in STS near $V_S$ = 370 mV, which depends on the tip and tunneling conditions. We observed that this charging peak shifts to higher energy ($\Delta V_r$ = 118 mV, defined with respect to the peak energy of an isolated atom, $\Delta V_r = V(r) - V(r > 10$ nm$)$) when the Co atom is near CuH$_H$ ($r$ = 3.8 nm) (Fig. 2).

In order to quantify this influence of the Cu donor, we measured the charging peak shift $\Delta V_r$ at various values of $r$ and for the three different Cu species (Fig. 3 / Supplementary Fig. 4). The studied Co-Cu pairs were created utilizing atomic manipulation of Co atoms. At distances of approximately $r$ > 10 nm, there is no shift of the ionization energy, i.e. $\Delta V_r$ is negligible. At smaller distances, we observe a monotonously increasing, strongly non-linear trend in $\Delta V_r$. The observed charging peak shift corresponds to a shift of the Co energy level that is being swept through $E_F$, as depicted in the inset of Fig. 3. This can stem from three effects: (1) Local band bending (i.e. gating of the conduction band) induced by the positively charged Cu atom. In the simple limit that the Cu atom can be approximated as a point charge, the generated electric field is not fully screened. Therefore, we can approximate this band bending by a Yukawa potential due to the screening from the substrate: $V_{Cu} = \frac{g}{r} e^{\frac{-r}{r_c}}$. Here, $g$ is a scaling constant and $r_c$ the effective screening length. (2) Changes in the tip-induced band-bending (TIBB), that additionally contributes when another mechanism (e.g. point 1) shifts the overall charging peak onset. We can assume that the TIBB scales linearly with applied bias in the relevant energy range (see for example [36]). Such linear changes cannot explain the non-linear behavior observed in Fig. 3, but it may change the scaling constant of the Yukawa potential. Furthermore, TIBB is strongly tip dependent. To account for these differences, we use $\Delta V_r$ instead of $V_S$ and the measurement was performed using different tips. As we show in Fig. 3, $\Delta V_r$ is roughly the same independent of the given tip. Therefore, we expect the variations in TIBB in the applied voltage range to be negligible. (3) DFT calculations reveal that the Co atom has an electric dipole moment ($\boldsymbol{p}_i$), where *i* labels the orbital state (see Supplementary Fig. 6). This calculated moment is oriented perpendicular to the surface. In the simplest limit, the subsequent change in energy equals $\Delta V_{\text{dipole}} = \boldsymbol{p} \cdot \boldsymbol{E}$. The electric field of the tip can lead to a strong change in this energy, but this does not depend on the distance of the Cu atom (*r*). The electric field of the tip does depend on the tip-sample separation and the bias $V_S$. The tip-sample separation is constant in the experiments, but the bias is not. Like in (2), the consequential changes to the experimentally probed $\Delta V_r$ are expected to be linear and cannot explain the non-linearity observed. As we detail in the supplemental information, this may explain why there are differences in the overall asymmetry as a function of applied voltage (as reported in our original paper [3]), compared to changing *r*. In the same way, the electric field of a Cu atom can couple to the dipole moment of the Co atom. However, since the field and dipole are nearly orthogonal, we can neglect this effect. Therefore, the only non-linear factor governing the shift of the charging peak energy level is point (1), the band bending induced by the presence of a Cu donor. Using this information, we fit the data in Fig. 3 with a Yukawa potential. The extracted effective screening length equals $r_c = 1.93$ nm.

To understand the influence of a local electric field on orbital memory, we studied the influence of a Cu-based donor on the stochastic noise of a nearby Co atom as a function of *r* (Fig. 4). By applying voltages typically above $V_S$ > 400 mV, isolated Co atoms exhibit telegraph noise resulting from stochastic switching between the bistable valencies: $Co_{high}$ and $Co_{low}$ (Fig. 4 (a)) [3]. We subsequently measured the telegraph noise of a Co atom, within the proximity of a Cu species. We measured the telegraph noise until we approach steady state (typically ~800 switching events), in order to be able to extract the state dependent lifetime ($\tau_{high}$ and $\tau_{low}$), as done in refs [3, 37]. We extracted $\tau_{high}$ and $\tau_{low}$ for multiple tips and atoms, with varying values of *r* (Fig. 4 (b-c)) at two sample biases: $V_S$ = 500 and 550 mV (more biases are shown in Supplementary Fig. 5). As in Fig. 3, the results in Fig. 4 (b-c) include Co-Cu$_T$, Co-Cu$_H$ and Co-CuH$_H$ pairs in various orientations and directions. The most striking feature is that the lifetime of the Co$_{high}$ state ($\tau_{high}$) is dramatically decreased by the proximity of the Cu donor, whereas the Co$_{low}$ state is only weakly perturbed in comparison. At $V_S$ = 500 mV, lifetime $\tau_{high}$ decreases from approximately 170 ms to 30 ms for decreasing $r$ from ~16 to 4 nm and at $V_S$ =550 mV, $\tau_{high}$ decreases more than an order of magnitude, from roughly 70 ms to 5 ms. We propose that the dependence of $\tau_{high}/\tau_{low}$ on $r$, is derived purely from the gating effects of the Cu – a local shifting of the bands. The reason why $\tau_{low}$ is not affected in proximity to Cu, is that the effective screening length of the given orbital state is roughly 2 nm (derived from the Yukawa fit in Fig. 3), thus the Cu is only a weak perturbation at $r$ > 4 nm. As we know from [3], the dielectric screening of the Co$_{high}$ state from the substrate is weaker, meaning the effective screening length should increase. Therefore, the onset of changes to $\tau_{high}$ should occur at larger values of $r$ compared to $\tau_{low}$.

It is interesting to compare the effect of the electric field of a Cu donor to the effect of the tip electric field on the stochastic switching of a Co atom. A higher bias $V_S$ presents a larger net electric field from the tip similar to Cu, and if the effect of a higher tip field is comparable to that of the field generated by a Cu atom, we should observe similar trends in both the mean lifetime and asymmetry of the orbital states. In the case of a higher applied bias voltage for an isolated atom without a Cu atom in the vicinity, we most prominently observe a strong decrease in the mean lifetime of the Co atom [3], whereas the influence of the gating field of Cu is strongly state selective, mainly influencing the lifetime of one orbital state (Co$_{high}$). This illustrates that there are different mechanisms at play between these two effects. As the tip field is most likely aligned with the dipole moment of Co, we conclude that the mechanism at play here likely results from dipolar coupling (i.e. a Stark-like effect). In this picture, a higher applied bias would strongly increase the energy in the system and mimic an effective temperature. In contrast, the Cu donor strongly influences the local band bending and therefore locally gates the orbital states of individual Co atom.

In conclusion, we demonstrated the effect of a controlled electric field generated by an individual Cu donor on the atomic orbital memory of Co on BP. Using both STM/STS and DFT calculations, we quantified the distance-dependent influence of individual Cu donors on the ionization energy of Co, as well as on the state-dependent lifetime in the stochastic limit. The monotonous increase in the charging energy of $Co_{low}$ with decreasing $r$ maps the downward band bending associated with the single Cu atoms and follows the characteristic behavior for a screened Coulomb interaction. We also find that the proximity to a local donor strongly influences the lifetime of the $Co_{high}$ state in the stochastic limit, whereas there is no influence on the lifetime of the $Co_{low}$ state. We attribute this to a gating effect of the Cu donor, impacting the valencies differently because of their distinct screening from the substrate. Notably, we detected no difference in the effect of the different Cu species on the ionization energy or stochastic behavior of Co. Furthermore, DFT calculations provided evidence of a state-dependent electric dipole moment for $Co_{low}$ and $Co_{high}$, which can explain the response of the lifetimes as a function of bias $V_S$ by a dipolar coupling between the Co atom and the tip electric field. Our findings illustrate how the state favorability of an atomic orbital memory, i.e. the energy landscape, can be tuned by an external electric field, analogous to a magnetic field in a spin-based memory. It remains to be seen how the spin states and crystal field are locally affected by the presence of the electric field.

**Acknowledgements**

The experimental part of this project was supported by the European Research Council (ERC) under the European Union's Horizon 2020 research and innovation programme (grant no. 818399). E. J. K. and A. A. K. acknowledge support from the NWO-VIDI project 'Manipulating the interplay between superconductivity and chiral magnetism at the single-atom level' with project no. 680-47-534. B. K. acknowledges the NWO-VENI project 'Controlling magnetism of single atoms on black phosphorus' with project no. 016.Veni.192.168. For the theoretical part of this work, A. N. R. and M. I. K. received funding from the European Research Council via Synergy Grant 854843 – FASTCORR.

**References**

1. T. O. Wehling, A. I. Lichtenstein and M. I. Katsnelson, Physical Review B **84** (23), 235110 (2011).
2. A. N. Rudenko, F. J. Keil, M. I. Katsnelson and A. I. Lichtenstein, Physical Review B **86** (7), 075422 (2012).
3. B. Kiraly, A. N. Rudenko, W. M. J. van Weerdenburg, D. Wegner, M. I. Katsnelson and A. A. Khajetoorians, Nature Communications **9** (1), 3904 (2018).
4. A. Kolmus, M. I. Katsnelson, A. A. Khajetoorians and H. J. Kappen, New Journal of Physics **22** (2), 023038 (2020).
5. U. Kamber, A. Bergman, A. Eich, D. Iuşan, M. Steinbrecher, N. Hauptmann, L. Nordström, M. I. Katsnelson, D. Wegner, O. Eriksson and A. A. Khajetoorians, Science **368** (6494), eaay6757 (2020).
6. B. Kiraly, E. J. Knol, W. M. J. van Weerdenburg, H. J. Kappen and A. A. Khajetoorians, Nature Nanotechnology **16** (4), 414-420 (2021).
7. F. Donati, S. Rusponi, S. Stepanow, C. Wäckerlin, A. Singha, L. Persichetti, R. Baltic, K. Diller, F. Patthey, E. Fernandes, J. Dreiser, Ž. Šljivančanin, K. Kummer, C. Nistor, P. Gambardella and H. Brune, Science **352** (6283), 318 (2016).
8. F. E. Kalff, M. P. Rebergen, E. Fahrenfort, J. Girovsky, R. Toskovic, J. L. Lado, J. Fernández-Rossier and A. F. Otte, Nature Nanotechnology **11** (11), 926-929 (2016).
9. F. D. Natterer, K. Yang, W. Paul, P. Willke, T. Choi, T. Greber, A. J. Heinrich and C. P. Lutz, Nature **543** (7644), 226-228 (2017).
10. J. Kim, S. S. Baik, S. H. Ryu, Y. Sohn, S. Park, B.-G. Park, J. Denlinger, Y. Yi, H. J. Choi and K. S. Kim, Science **349** (6249), 723 (2015).
11. S.-W. Kim, H. Jung, H.-J. Kim, J.-H. Choi, S.-H. Wei and J.-H. Cho, Physical Review B **96** (7), 075416 (2017).
12. C. Han, Z. Hu, L. C. Gomes, Y. Bao, A. Carvalho, S. J. R. Tan, B. Lei, D. Xiang, J. Wu, D. Qi, L. Wang, F. Huo, W. Huang, K. P. Loh and W. Chen, Nano Letters **17** (7), 4122-4129 (2017).
13. B. Kiraly, E. J. Knol, K. Volckaert, D. Biswas, A. N. Rudenko, D. A. Prishchenko, V. G. Mazurenko, M. I. Katsnelson, P. Hofmann, D. Wegner and A. A. Khajetoorians, Physical Review Letters **123** (21), 216403 (2019).
14. See Supplementary Information for methods and detailed information on: Cu species, Cu doping of BP, the ionization energy of Co atoms, bias dependence of stochastic switching of Co, and calculations of the electric dipole moment of Co. This includes Refs. [15-27].
15. P. E. Blöchl, Physical Review B **50** (24), 17953-17979 (1994).
16. G. Kresse and J. Furthmüller, Physical Review B **54** (16), 11169-11186 (1996).
17. G. Kresse and D. Joubert, Physical Review B **59** (3), 1758-1775 (1999).
18. J. P. Perdew, K. Burke and M. Ernzerhof, Physical Review Letters **77** (18), 3865-3868 (1996).
19. A. Brown and S. Rundqvist, Acta Crystallographica **19** (4), 684-685 (1965).
20. N. Marzari, A. A. Mostofi, J. R. Yates, I. Souza and D. Vanderbilt, Reviews of Modern Physics **84** (4), 1419-1475 (2012).
21. A. A. Mostofi, J. R. Yates, Y.-S. Lee, I. Souza, D. Vanderbilt and N. Marzari, Computer Physics Communications **178** (9), 685-699 (2008).
22. J. Tersoff and D. R. Hamann, Physical Review B **31** (2), 805-813 (1985).
23. A. Zhao, Q. Li, L. Chen, H. Xiang, W. Wang, S. Pan, B. Wang, X. Xiao, J. Yang, J. G. Hou and Q. Zhu, Science **309** (5740), 1542 (2005).
24. N. Baadji, S. Kuck, J. Brede, G. Hoffmann, R. Wiesendanger and S. Sanvito, Physical Review B **82** (11), 115447 (2010).
25. B. W. Heinrich, G. Ahmadi, V. L. Müller, L. Braun, J. I. Pascual and K. J. Franke, Nano Letters **13** (10), 4840-4843 (2013).
26. A. A. Khajetoorians, T. Schlenk, B. Schweflinghaus, M. dos Santos Dias, M. Steinbrecher, M. Bouhassoune, S. Lounis, J. Wiebe and R. Wiesendanger, Physical Review Letters **111** (15), 157204 (2013).
27. V. C. Zoldan, R. Faccio and A. A. Pasa, Scientific Reports **5** (1), 8350 (2015).
28. B. Kiraly, N. Hauptmann, A. N. Rudenko, M. I. Katsnelson and A. A. Khajetoorians, Nano Letters **17** (6), 3607-3612 (2017).
29. S. P. Koenig, R. A. Doganov, L. Seixas, A. Carvalho, J. Y. Tan, K. Watanabe, T. Taniguchi, N. Yakovlev, A. H. Castro Neto and B. Özyilmaz, Nano Letters **16** (4), 2145-2151 (2016).
30. S. W. Lee, L. Qiu, J. C. Yoon, Y. Kim, D. Li, I. Oh, G.-H. Lee, J.-W. Yoo, H.-J. Shin, F. Ding and Z. Lee, Nano Letters (2021).
31. Z. Lin, J. Wang, X. Guo, J. Chen, C. Xu, M. Liu, B. Liu, Y. Zhu and Y. Chai, InfoMat **1** (2), 242-250 (2019).

32. Y. Zheng, H. Yang, C. Han and H. Y. Mao, Advanced Materials Interfaces **7** (17), 2000701 (2020).
33. R. M. Feenstra, Journal of Vacuum Science & Technology B: Microelectronics and Nanometer Structures Processing, Measurement, and Phenomena **21** (5), 2080-2088 (2003).
34. F. Marczinowski, J. Wiebe, F. Meier, K. Hashimoto and R. Wiesendanger, Physical Review B **77** (11), 115318 (2008).
35. K. Teichmann, M. Wenderoth, S. Loth, R. G. Ulbrich, J. K. Garleff, A. P. Wijnheijmer and P. M. Koenraad, Physical Review Letters **101** (7), 076103 (2008).
36. G. J. de Raad, D. M. Bruls, P. M. Koenraad and J. H. Wolter, Physical Review B **66** (19), 195306 (2002).
37. A. A. Khajetoorians, B. Baxevanis, C. Hübner, T. Schlenk, S. Krause, T. O. Wehling, S. Lounis, A. Lichtenstein, D. Pfannkuche, J. Wiebe and R. Wiesendanger, Science **339** (6115), 55 (2013).

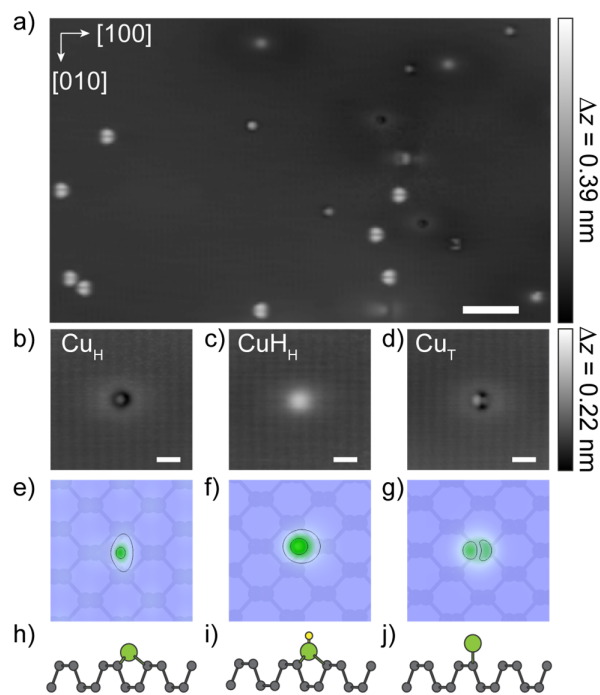

**Figure 1**. (a) STM image of individual Co and Cu atoms on the surface of BP. ($V_S$ = -400 mV, $I_t$ = 20 pA, scalebar = 5 nm). (b-d) STM images of the charge density of three Cu species: (b) Cu residing in a hollow site ($Cu_H$), (c) hydrogenated Cu residing in a hollow site ($CuH_H$) and (d) Cu residing on a top site ($Cu_T$). ($V_S$ = -400 mV, $I_t$ = 60 pA, scalebar = 1 nm). (e-g) *Ab initio* calculations of the relaxed charge density of (e) a $Cu_H$ atom, (f) $CuH_H$ and (g) a $Cu_T$ atom. (h-j) Schematics of the relaxed atomic adsorption geometries of (h) $Cu_H$, (i) $CuH_H$ and (j) $Cu_T$.

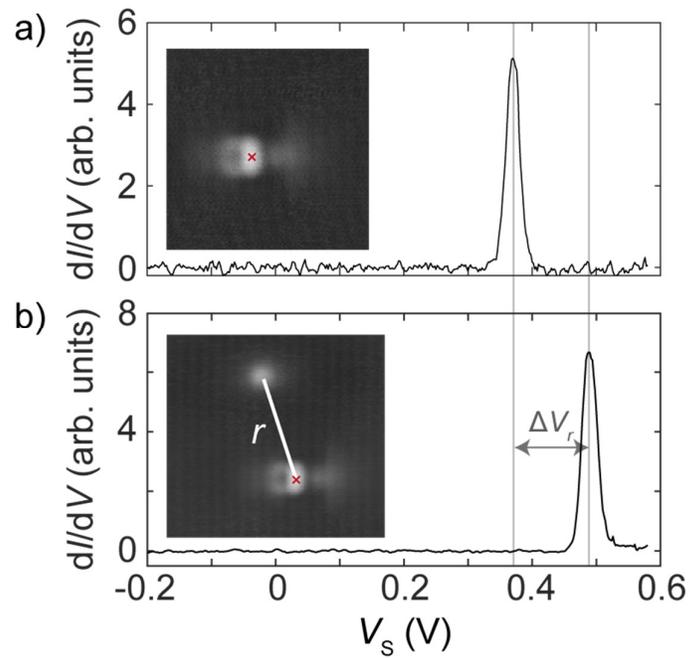

**Figure 2.** (a) d$I$/d$V$ spectrum taken on an isolated Co$_{low}$ atom (distance to nearest Cu species $r > 10$ nm). Inset: STM image isolated Co$_{low}$ atom ($V_S$ = -400 mV, $I_t$ = 60 pA). (b) d$I$/d$V$ spectrum taken on a Co$_{low}$ atom in the vicinity of CuH$_H$ (distance $r$ = 3.8 nm). The shift of the ionization peak with respect to the isolated atom is indicated by $\Delta V_r$ and equals 118 mV. Inset: STM image of the Co$_{low}$ - CuH$_H$ pair. The red X in both insets (a-b) mark the locations where the d$I$/d$V$ spectra were taken. ($V_S$ = -400 mV, $I_t$ = 60 pA).

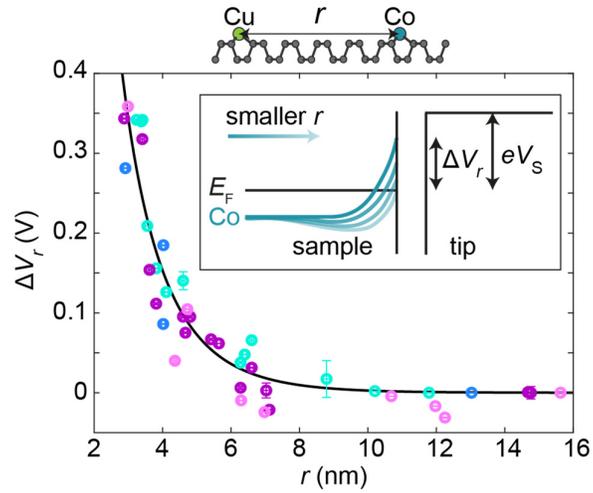

**Figure 3.** Shift of the ionization peak energy $\Delta V_r$ of single Co$_{low}$ atoms as a function of distance to a Cu species $r$. The shift is defined with respect to the peak energy of an isolated atom: $\Delta V_r = V(r) - V(r > 10$ nm$)$. Different colors represent different microtips. The solid line is a fit to the Yukawa potential, with parameters $g = 4.88$; $r_c = 1.93$ nm. In the inset, a schematic band diagram shows downward band bending of the Co energy level (which is pulled through $E_F$ by TIBB) in the presence of a Cu species.

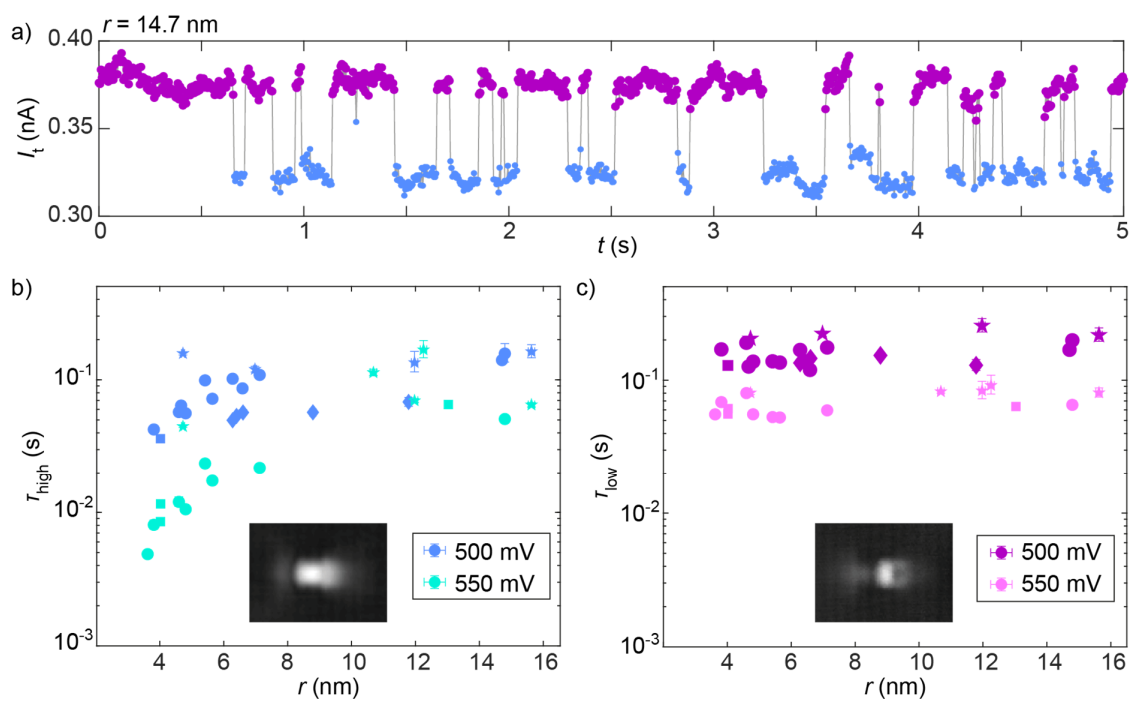

**Figure 4**. (a) Two-state telegraph noise signal of an isolated Co atom (nearest Cu donor at $r$ = 14.7 nm), with $Co_{low}$ in purple and $Co_{high}$ in blue. (b-c) State lifetime $\tau$ of (b) $Co_{high}$ and (c) $Co_{low}$ as a function of $r$. Different symbols represent different microtips and different colors represent different biases (500 mV and 550 mV). In the insets, STM images of the corresponding states $Co_{high}$ and $Co_{low}$ are displayed ($V_S$ = -400 mV, $I_t$ = 60 pA).

# Supplementary Information

**Table of contents**

# Methods

*Scanning Tunneling Microscopy/Spectroscopy*

STM and STS measurements were performed with a commercial Omicron low-temperature STM in ultrahigh vacuum ($p < 1 \times 10^{-10}$ mbar) and at a base temperature of $T = 4.4$ K. The bias was applied to the sample ($V_S$). All measurements were performed with electrochemically etched W tips. The tips were treated *in situ* by electron bombardment and field emission, whereafter they were dipped and characterized on a clean Au(111) surface. The STM images in this work were all acquired using constant-current feedback. STS measurements were performed using a lock-in technique to directly measure $dI/dV$. A modulation was applied to the bias signal of $V_{mod} = 6$ mV at a frequency of $f_{mod} = 817.7$ Hz, unless stated otherwise. All telegraph measurements were acquired with the tip at constant height. The tip height was stabilized with constant-current feedback on the bare BP at $I_t = 20$ pA, $V_S = -400$ mV, before opening the feedback loop. BP crystals were purchased from HQ graphene. The crystals were cleaved with scotch tape in UHV and immediately transferred to the microscope for *in situ* characterization. Co and Cu were evaporated directly into the microscope with $T_{sample} < 5$ K during the dosing procedure. Atomic manipulation of the Co atoms was performed by dragging the top-site atoms along the [010] direction in constant-current mode with $-130$ mV $< V_S < -100$ mV and 6 nA $< I_t <$ 12 nA. After a Co atom reached a desirable location (with respect to a Cu donor), it was manipulated into a hollow site by means of a voltage pulse of $V_S < -700$ mV.

*Density Functional Theory*

DFT calculations were carried out using the projected augmented-wave method (PAW) [1] as implemented in the Vienna *ab initio* simulation package (VASP) [2, 3]. Exchange and correlation effects were taken into account within the spin-polarized generalized gradient approximation (GGA-PBE) [4]. The Hubbard-U correction was not applied to Cu atoms because the 3*d* shell of Cu is almost entirely filled and, therefore, is not sensitive to additional local Coulomb interaction. An energy cutoff of 300 eV for the plane-wave basis and the convergence threshold of $10^{-6}$ eV were used, which is sufficient to obtain numerical accuracy. Pseudopotentials were taken to include 3*s* and 3*p* valence electrons for P atom, as well as 3*d* and 4*s* valence electrons for Cu atoms. The BP surface was modeled in the slab geometry by a single BP layer with dimensions (3*a* × 4*b*) ≈ (13.1 × 13.3) Å with atomic positions fixed to the experimental parameters of bulk BP [5]. Vertical separation between the layers was set to 20 Å. The Brillouin zone was sampled by a uniform distribution of ***k***-points on a (8 × 8) mesh. The position of the Cu atom was relaxed considering top and hollow surface sites as starting points. The projection of the electronic bands on specific atomic states was performed using the formalism of Wannier functions [6] implemented in the wannier90 package [7].

The spatial charge density distributions shown in Fig. 1e-g were calculated by performing band decomposition of the total DFT charge densities and averaging them over the energy interval of ~0.3 eV in the valence band edge, similar to [8]. The resulting charge distribution reflects the surface charge densities typical to the low-energy hole states in BP with Cu adatoms and, therefore, can be associated with the experimental STM images shown in Fig. 1b-d. This procedure is closely related to the Tersoff-Hamann scheme [9]. Possible mismatch between the simulated and experimental STM images can be related to the effect of the STM tip, which is not explicitly considered in our approach.

The charge transfer from Cu to BP, Δ*n*, was estimated as a difference between the number of Cu valence electrons in a free atom (*N*=11) and in the atom adsorbed on BP (*n*), that is Δ*n* = *N* − *n*, where *n* is calculated by integrating the density of states *g(E)* projected onto the *s*- and *d*-orbitals of Cu, $n = \int_{-\infty}^{E_F} g(E)dE$ with *E*$_F$ being the Fermi energy.

The electric dipole moment at finite electric fields was calculated from the vertical distribution of the charge density $n(z) = \frac{1}{S}\int n(\mathbf{r})dxdy$ by applying a saw-tooth potential in the direction normal to the BP surface. The calculations for a Co adatom on BP were performed for two orbital configurations (low-spin and high-spin) using the same parameters as in ref [8].

### Detailed information about Cu species

Based on a binding site analysis comparing the charge density in constant-current imaging with the underlying BP lattice and *ab initio* calculations of the relaxed charge density (Fig. 1), we found that two of three observed Cu species reside in a hollow site, and one species resides on a top site. Both hollow site species are identified as isotropic protrusions and can be distinguished by the presence/absence of a circular depression directly around the center of this protrusion and distinct apparent heights of 13 ± 1 pm / 97 ± 2 pm at $V_S$ = -400 mV for the species in Fig. 1 (b) and (c) respectively.

The Cu atoms residing in the top site ($Cu_T$) are identified by their bi-lobed shapes and an apparent height of 52 ± 4 pm at $V_S$ = -400 mV. Two types of these top site atoms are observed, related through mirror symmetry along the zig-zag [010] direction (Supplementary Fig. 1(b-c)). Similarly, mirror-symmetric species are also observed for the different Co species [8]. While both hollow site species can potentially be attributed to bistable valency, like for Co, we show that the difference between both Cu atoms at the hollow site results from hydrogen adsorption: in Supplementary Fig. 1(a), an image is displayed showing a (de)hydrogenation event of a Cu atom in the hollow site. Such an event occurs occasionally during an STM image and is in line with observations of hydrogenation of single atoms and molecules on various surfaces [10-14]. After deposition at $T$ < 5 K, ~68% of the Cu atoms reside in a hollow site and ~32% reside on a top site. The amount of hydrogenated hollow site species increased over time.

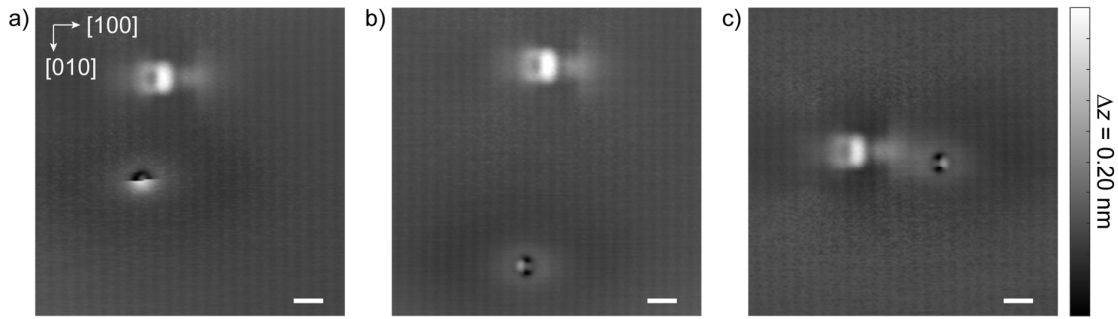

**Supplementary Figure 1**. (a) Hydrogenation event of a $Cu_H$ atom, during an STM image. (b-c) STM images of Co-$Cu_T$ pairs. The $Cu_T$ atoms are mirror-symmetric in the [010] direction. Parameters for (a-c): $V_s$ = -400 mV, $I_t$ = 60 pA, scalebar = 1 nm.

## Copper doping of black phosphorus

To examine the effect of Cu doping on the band structure of BP, scanning tunneling spectroscopy (STS) and density functional theory (DFT) calculations were performed. Both methods reveal that deposition of Cu on BP leads to *n*-doping, in line with literature [15-18]. Single Cu atoms were deposited on several BP samples, where after the shift in the bandgap of BP could be measured with STS at sufficiently high surface coverage. Supplementary Figure 2(a) shows d$I$/d$V$ spectra of pristine BP and Cu-doped BP at a coverage of $n_{Cu}$ = 3.5x10$^{12}$ cm$^{-2}$, revealing a clear shift of the bandgap to lower energy. Supplementary Figure 2(b) shows the band edges from multiple pristine and Cu-doped BP samples, extracted using the same method as in ref [19]. For each extracted bandgap, between 25 and 160 d$I$/d$V$ spectra were analyzed. The extracted band edges for pristine BP are in good agreement with refs [19, 20].

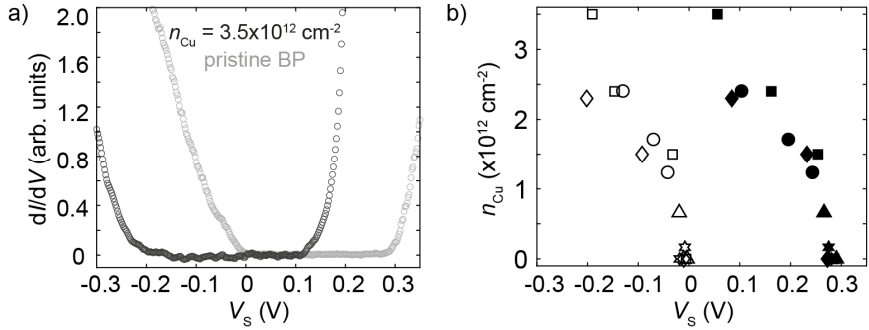

**Supplementary Figure 2**. (a) d$I$/d$V$ spectra for pristine BP (gray) and Cu-doped BP (black) at $n_{Cu}$ = 3.5x10$^{12}$ cm$^{-2}$. (b) Valence band maxima (empty markers) and conduction band minima (filled markers) extracted from multiple d$I$/d$V$ spectra taken in different spatial locations, as a function of the areal density of Cu ($n_{Cu}$). Each symbol represents iterations of depositing Cu on the same BP sample. The standard error of the extracted band edges is smaller than the symbol sizes. For all spectra in (a) and (b): $I_{stabilization}$ = 100 pA, $f$ = 817.7 Hz, $V_{mod}$ = 10 mV.

Supplementary Figure 3 shows the band structure calculated within DFT for a Cu atom on BP in the hollow (Cu$_H$) and top sites (Cu$_T$), and in the hollow site with hydrogenation (CuH$_H$). The color indicates the occupation of the 3$d$ and 4$s$ shell of Cu. In all cases, the phosphorus states as well as the band gap are easily distinguishable. In case of Cu$_H$ and CuH$_H$, the system is nonmagnetic and the electronic bands are degenerate in spin. For Cu$_H$ there is one band associated with Cu states at the Fermi energy, which is half-filled (Supplementary Fig. 3(a)). This band becomes fully occupied after hydrogenation (Supplementary Fig. 3(b)). The situation when Cu is adsorbed on the top site (Cu$_T$) is different (Supplementary Fig. 3(c-d)). In this case, the Cu states are non-spin-degenerate and Cu atoms carry a magnetic moment of ~0.5 μ$_B$. The corresponding Cu band becomes fully occupied for the spin-up states and fully empty for the spin-down states. Non-integer magnetic moment is the result of a hybridization between copper and phosphorus states.

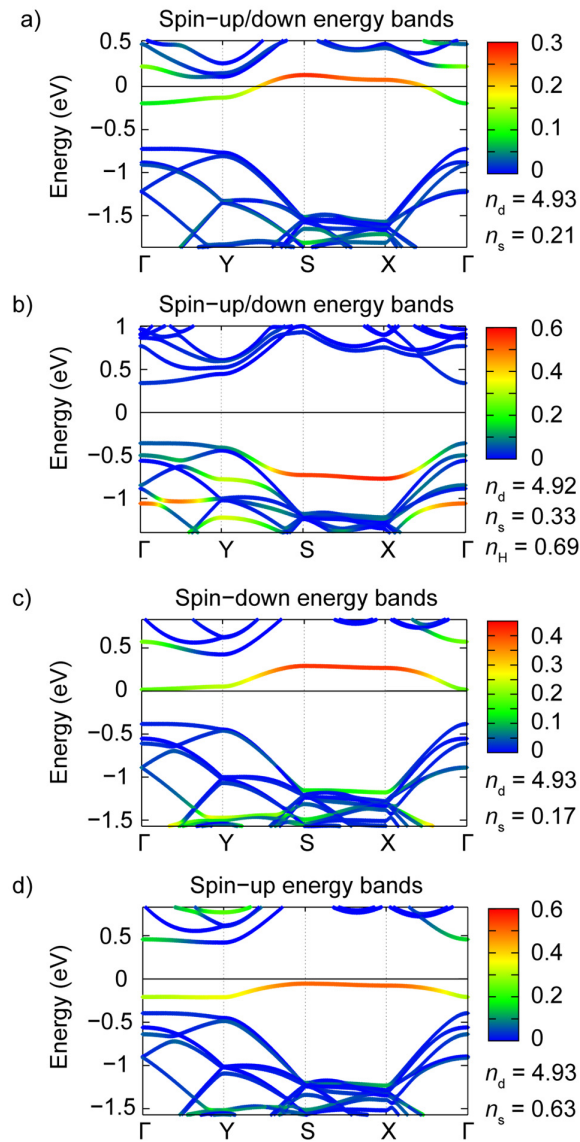

**Supplementary Figure 3.** Band structure of Cu-doped BP for (a) Cu$_H$, (b) CuH$_H$, (c-d) Cu$_T$. The color scale stands for the contribution of the different orbitals: blue bands stem exclusively from *p* orbitals, while red indicates a contribution from Cu orbitals. The occupation of the 3*d* and 4*s* shell of Cu are given below the colorbar.

## Ionization peak energy of Co$_{low}$ atoms

The ionization peak energy $V_{S,peak}$ of single Co$_{low}$ atoms was extracted using Gaussian fits of the peaks in the d$I$/d$V$ spectra. The results for different tips and different Cu species are shown in Supplementary Figure 4, as a function of distance $r$. For each microtip, the qualitative trend is the same: there is a monotonous increase of the ionization peak energy as $r$ decreases. Quantitatively, there is a different offset for each microtip. This can be explained by differences in the band bending condition for each microtip. To be able to compare the results from different microtips, in Figure 3 the energy difference $\Delta V_r$ is shown w.r.t. an isolated Co atom ($r > 10$ nm).

The results in Supplementary Figure 4 and Figure 3 include Co-Cu$_T$, Co-Cu$_H$ and Co-CuH$_H$ pairs in various directions with respect to the BP crystallographic axes (although most pairs were built roughly along the [100] and [010] directions) and both Co orientations with respect to the mirror-symmetry in the [100] direction. As seen in Supplementary Fig. 4, there is no difference in the trend for different Cu species. This indicates that the band bending profiles are similar for all Cu species and insensitive to the relative orientation between the atomic displacement vector ($r$) and the lattice vectors ([100] and [110]) of the BP.

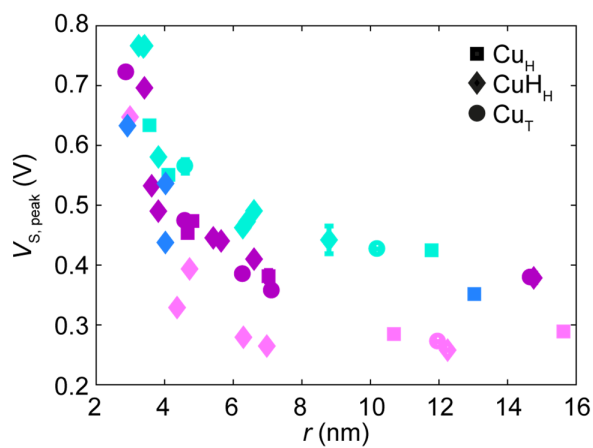

**Supplementary Figure 4**. Extracted ionization peak energy of single Co$_{low}$ atoms as a function of distance to a Cu species $r$. Different colors represent different microtips and correspond to the colors in Figure 3. The different markers indicate the Cu species that is closest to the measured Co atom.

## Bias dependence of lifetimes in stochastic regime

In Supplementary Figure 5, the lifetimes of the $Co_{high}$ and $Co_{low}$ states are shown as a function of Co-Cu distance *r* and for six different tip-sample biases $V_S$ between 425 mV and 600 mV. The observed trends at 500 mV and 550 mV (Fig. 4) are identical for the other probed tip-sample biases: $\tau_{high}$ decreases for decreasing *r* and $\tau_{low}$ remains approximately constant as a function of *r*. We note that at the lowest bias shown here (425 mV), there is not as much data so that the trend is less clear.

There is a difference in the response of $\tau_{low}$ and $\tau_{high}$ (measured in the ionized state) to the distant-dependent presence of a Cu atom, and changes in the bias $V_S$. That is, a Cu donor in the vicinity does not affect $\tau_{low}$ and $\tau_{high}$ in the same way, but the bias $V_S$ does: both lifetimes decrease for higher biases. As discussed in the main paper, we propose that the distance-dependence response to a Cu atom is caused by a local gating effect. Concerning the bias dependence, we propose that the dipolar coupling between Co and the tip electric field plays an important role, which is independent of *r*. We suggest that the changes in $\tau_{low}/\tau_{high}$ as a function of applied bias $V_S$, which was also investigated in [8], may come from a dipolar coupling between the tip electric field and a potential out-of-plane dipole moment in the Co atom. This is reminiscent of a Stark shift, where increasing the tip electric field can be viewed as an overall increase in the energy of the potential landscape (i.e. mimicking temperature in an Arrhenius model). We note that this picture neglects any in-plane orientation of the static or dynamic dipole moments of either Co state, meaning we neglect the effect of Cu-Co dipole coupling. This is a good assumption for the neutral Co atom, but for the ionized Co atom, further calculations would be needed.

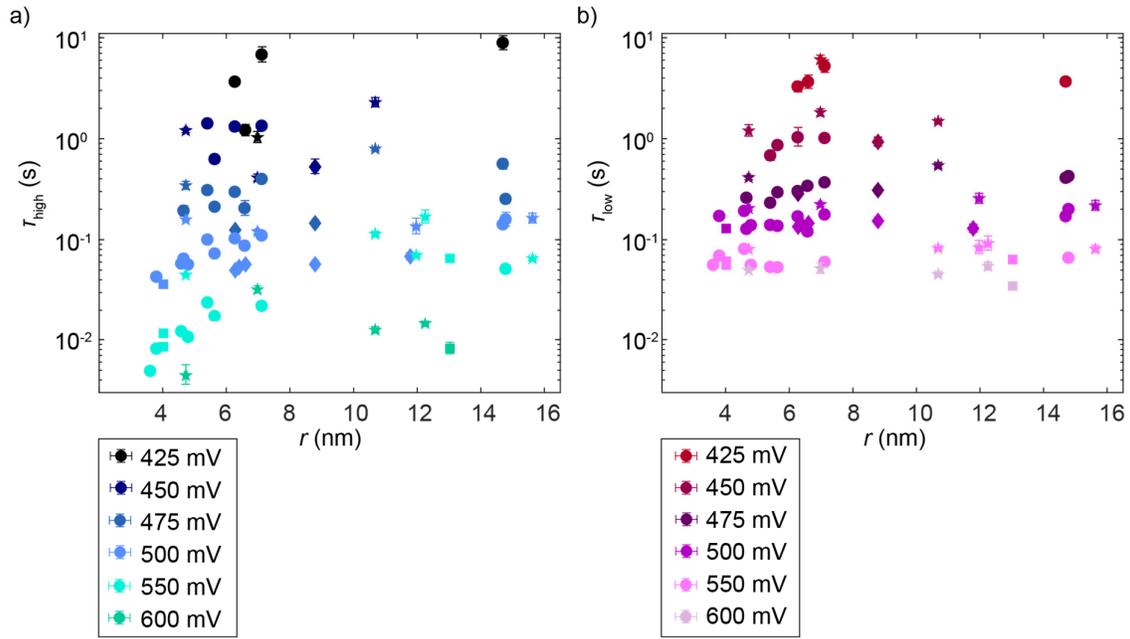

**Supplementary Figure 5**. (a-b) State lifetime $\tau$ of (a) Co$_{high}$ and (b) Co$_{low}$ as a function of Cu-Co distance $r$, for six different biases and several different microtips. Different symbols represent different microtips and different colors represent the different biases between 425 mV and 600 mV.

## Electric dipole moment of Co atoms

Supplementary Figure 6 shows (a) energy variation, (b) vertical distance variation and (c) electric dipole moment as a function of vertical electric field calculated within DFT for low-spin and high-spin states of a Co adatom on BP. The energy variation is quadratic in electric field, and is similar for both states of the adatom. The vertical distance variation is the change in height of the Co atom with respect to its equilibrium position, without an electric field. In contrast to the energy variation, the vertical distance changes differently for the low- and high-spin states. While the former demonstrates a linear dependence, asymmetric with respect to the sign of electric field, the latter resembles a quadratic dependence. This behavior indicates that the vertical charge distribution is essentially different for the two states. This is further confirmed by the calculations of the electric dipole moment shown in Supplementary Fig. 6(c). Although the overall dependence on the electric field is similar, the absolute values of the dipole moment at zero electric field are significantly different, resulting in -0.12 and -0.22 $e$Å for the low- and high-spin states, respectively.

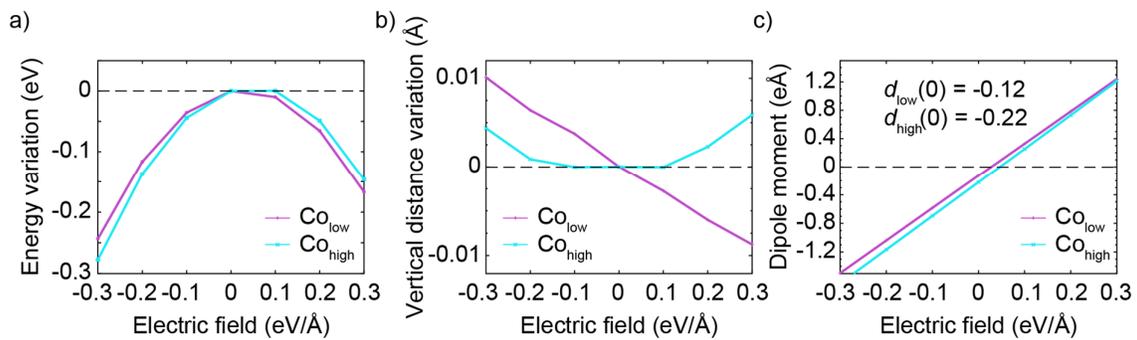

**Supplementary Figure 6**. Ab initio calculations of (a) energy variation, (b) vertical distance variation and (c) dipole moment as a function of vertical electric field shown for low-spin and high-spin states of Co adatom on BP.

# References

1. P. E. Blochl, Physical Review B **50** (24), 17953-17979 (1994).
2. G. Kresse and J. Furthmuller, Physical Review B **54** (16), 11169-11186 (1996).
3. G. Kresse and D. Joubert, Physical Review B **59** (3), 1758-1775 (1999).
4. J. P. Perdew, K. Burke and M. Ernzerhof, Physical Review Letters **77** (18), 3865-3868 (1996).
5. A. Brown and S. Rundqvist, Acta Crystallographica **19** (4), 684-685 (1965).
6. N. Marzari, A. A. Mostofi, J. R. Yates, I. Souza and D. Vanderbilt, Reviews of Modern Physics **84** (4), 1419-1475 (2012).
7. A. A. Mostofi, J. R. Yates, Y.-S. Lee, I. Souza, D. Vanderbilt and N. Marzari, Computer Physics Communications **178** (9), 685-699 (2008).
8. B. Kiraly, A. N. Rudenko, W. M. J. van Weerdenburg, D. Wegner, M. I. Katsnelson and A. A. Khajetoorians, Nature Communications **9** (1), 3904 (2018).
9. J. Tersoff and D. R. Hamann, Physical Review B **31** (2), 805-813 (1985).
10. A. Zhao, Q. Li, L. Chen, H. Xiang, W. Wang, S. Pan, B. Wang, X. Xiao, J. Yang, J. G. Hou and Q. Zhu, Science **309** (5740), 1542 (2005).
11. N. Baadji, S. Kuck, J. Brede, G. Hoffmann, R. Wiesendanger and S. Sanvito, Physical Review B **82** (11), 115447 (2010).
12. B. W. Heinrich, G. Ahmadi, V. L. Müller, L. Braun, J. I. Pascual and K. J. Franke, Nano Letters **13** (10), 4840-4843 (2013).
13. A. A. Khajetoorians, T. Schlenk, B. Schweflinghaus, M. dos Santos Dias, M. Steinbrecher, M. Bouhassoune, S. Lounis, J. Wiebe and R. Wiesendanger, Physical Review Letters **111** (15), 157204 (2013).
14. V. C. Zoldan, R. Faccio and A. A. Pasa, Scientific Reports **5** (1), 8350 (2015).
15. S. P. Koenig, R. A. Doganov, L. Seixas, A. Carvalho, J. Y. Tan, K. Watanabe, T. Taniguchi, N. Yakovlev, A. H. Castro Neto and B. Özyilmaz, Nano Letters **16** (4), 2145-2151 (2016).
16. S. W. Lee, L. Qiu, J. C. Yoon, Y. Kim, D. Li, I. Oh, G.-H. Lee, J.-W. Yoo, H.-J. Shin, F. Ding and Z. Lee, Nano Letters (2021).
17. Z. Lin, J. Wang, X. Guo, J. Chen, C. Xu, M. Liu, B. Liu, Y. Zhu and Y. Chai, InfoMat **1** (2), 242-250 (2019).
18. Y. Zheng, H. Yang, C. Han and H. Y. Mao, Advanced Materials Interfaces **7** (17), 2000701 (2020).
19. B. Kiraly, N. Hauptmann, A. N. Rudenko, M. I. Katsnelson and A. A. Khajetoorians, Nano Letters **17** (6), 3607-3612 (2017).
20. B. Kiraly, E. J. Knol, K. Volckaert, D. Biswas, A. N. Rudenko, D. A. Prishchenko, V. G. Mazurenko, M. I. Katsnelson, P. Hofmann, D. Wegner and A. A. Khajetoorians, Physical Review Letters **123** (21), 216403 (2019).